\documentclass[12pt]{iopart}
\usepackage{iopams}
\usepackage{dsfont}
\usepackage[dvips]{graphicx}
\newcommand{\half}{\mbox{$\textstyle \frac{1}{2}$}}

\newcommand{\re}{\mbox{$\rm e$}}
\newcommand{\ri}{\mbox{$\rm i$}}
\newcommand{\rd}{\mbox{$\rm d$}}

\begin{document}
\newtheorem{prop}{Proposition}

\title[Hamiltonian statistical mechanics]
{Hamiltonian statistical mechanics}

\author[D~C~Brody, D~C~P~Ellis, and D~D~Holm]
{Dorje~C~Brody, David~C~P~Ellis, and Darryl~D~Holm}

\address{Department of Mathematics, Imperial College London,
London SW7 2AZ, UK}

\begin{abstract}
A framework for statistical-mechanical analysis of quantum
Hamiltonians is introduced. The approach is based upon a gradient
flow equation in the space of Hamiltonians such that the
eigenvectors of the initial Hamiltonian evolve toward those of the
reference Hamiltonian. The nonlinear double-bracket equation
governing the flow is such that the eigenvalues of the initial
Hamiltonian remain unperturbed. The space of Hamiltonians is
foliated by compact invariant subspaces, which permits the
construction of statistical distributions over the Hamiltonians. In
two dimensions, an explicit dynamical model is introduced, wherein
the density function on the space of Hamiltonians approaches an
equilibrium state characterised by the canonical ensemble. This is
used to compute quenched and annealed averages of quantum
observables.
\end{abstract}

\submitto{\JPA}
%
\vspace{0.4cm}

In the conventional approach to statistical mechanics the
Hamiltonian of the system under consideration is held fixed. If the
system is in equilibrium with a heat bath, then uncertainties in the
state of the system arise from `thermal noise' due to random
interactions with the bath. The equilibrium distribution over the
state space of the system (configuration space of a classical spin
system, classical phase space, or the space of pure quantum states)
is then established. However, in some cases---as in amorphous
alloys---the Hamiltonian need not be fixed, and  may even fluctuate
owing to thermal or other intrinsic sources. Observable effects
arising from such Hamiltonians may even be significant in the
quantum domain.

The purpose of the present paper is to introduce a theoretical
framework for an equilibrium theory of Hamiltonians. The fact that
parameters or matrix elements of the Hamiltonian themselves are
subject to random fluctuations for some systems has long been
recognised in the literature of spin glass~\cite{Virasoro} or random
matrix theory~\cite{RM}. The novel idea introduced here, as
distinguished from that considered in the theory of spin glass or
random matrices, is the construction of equilibrium distributions
over \textit{invariant subspaces of the space of quantum
Hamiltonians} by using a gradient flow equation on the space of
Hermitian matrices.

In classical statistical mechanics the notion of a gradient flow
plays an important role in describing the approach to equilibrium: A
system immersed in a heat bath naturally tends to release its energy
into the environment and thus approach its minimum energy state, and
this tendency is characterised by a Hamiltonian gradient flow. An
equilibrium state is attained when this flow is on the average
counterbalanced by thermal noise, where the magnitude of the noise
is determined by the temperature of the bath. Accordingly, we shall
introduce a \textit{gradient flow equation} on the space of
Hamiltonians such that the eigenstates of an arbitrary initial
Hamiltonian $H_0$ at time $t=0$ tend toward alignment with those of
a reference Hamiltonian, denoted by $G$. Thus, $G$ plays the role of
the `fixed' Hamiltonian in conventional quantum statistical
mechanics. The eigenstates of $H_t$ thus evolve toward those of $G$
under the flow. By introducing of a suitable noise term, we then
characterise the approach to an equilibrium distribution.

The paper is organised as follows. The key results concerning the
properties of the double-bracket equation that generates the
gradient flow are summarised first in the Proposition. The notion of
a double-bracket flow was first introduced by Landau and Lifshitz in
the context of characterising dispersions in magnetism
\cite{landau}. In its `modern form' it was introduced by Brockett
\cite{brockett} and has been successfully applied to many areas,
such as optimal control, linear programming, sorting algorithms, and
dissipative systems. Although some assertions of the Proposition are
valid in all dimensions, we shall analyse only the two-dimensional
case in full detail. We subsequently construct an explicit model for
the `equilibrisation' of $2\times2$ quantum Hamiltonians, such that
the stationary state is given by the canonical distribution. The
resulting statistical theory of quantum Hamiltonians can be extended
to a modification of quantum statistical mechanics. In particular,
we work out the quenched and annealed averages of quantum
observables. We conclude by indicating how the analysis can be
extended to higher dimensions.

\textbf{Proposition}. \textit{Let $H_t$ and $G$ be arbitrary
$2\times2$ Hermitian matrices, where $H_t$ is time dependent and $G$
is fixed. Let $H_t$ satisfy the double-bracket evolution equation
\begin{eqnarray}
\frac{\rd H_t}{\rd t} = -\lambda \left[H_t,[H_t,G]\right] \qquad
(\lambda\in{\mathds R}_+), \label{eq:1}
\end{eqnarray}
with initial condition $H_0$. Then the evolution {\rm (\ref{eq:1})}
is isospectral, i.e. the eigenvalues of $H_0$ are preserved under
{\rm (\ref{eq:1})}, and $\lim_{t\to\infty} [H_t,G]=0$. Furthermore,
the space of Hermitian Hamiltonians is foliated by a family of
invariant 2-spheres ${\mathcal L}$, and {\rm (\ref{eq:1})} induces a
gradient flow on each ${\mathcal L}$.}

We remark that in terms of the Hermitian operator $X=\ri [H,G]$ the
double-bracket evolution (\ref{eq:1}) can be rewritten as $\rd H =
\ri\lambda[H,X]\rd t$, which formally is just the Heisenberg
equation of motion. However, owing to  the $H$-dependence of $X$ the
evolution is nonunitary. We also note that in units $\hbar=1$ the
parameter $\lambda$ has dimension $[{\rm Energy}]^{-1}$. The
Hamiltonians $H_0$ and $G$ are both assumed nondegenerate;
otherwise, if at least one of the Hamiltonians is degenerate, then
$H_0$ is a fixed point of the flow. We now proceed to establish the
Proposition.

The fact that equation (\ref{eq:1}) asymptotically drives $H_t$
toward $[H_t,G]=0$, \textit{irrespective of the dimensionality of
the matrices}, follows from the relation
\begin{eqnarray}
\frac{\rd}{\rd t}\,{\tr}\left(H_t-G \right)^2 = -2{\tr} \left([G,
H_t]^\dagger [G, H_t] \right) \leq 0 , \label{eq:9}
\end{eqnarray}
where the equality is attained if and only if $[H_t,G]=0$. To see
that (\ref{eq:1}) defines an isospectral flow (which is also valid
irrespective of the dimensionality of the matrices) we note that the
right side of (\ref{eq:1}) can be written in the form $\lambda
\rd(\re^{-{\rm i}sX} H_t \re^{{\rm i}sX})/\rd s|_{s=0}$. The
isospectral property then follows from the relation $\det(
\re^{-{\rm i}sX} H_t \re^{{\rm i}sX} - E{\mathds 1})=\det(H_t-E
{\mathds 1})$. To prove that the orbit of the flow for a given
initial value $H_0$ lies on a two-sphere ${\mathcal L}$ (which is
isomorphic to the space of pure states for a two-level system), and
that (\ref{eq:1}) defines a gradient flow on ${\mathcal L}$, we
shall solve (\ref{eq:1}) explicitly for the case of $2\times2$
Hermitian matrices.

Let the $2\times2$ Hamiltonian $H_t$ be represented in terms of the
Pauli matrices as
\begin{eqnarray}
H_t=\half\,u_t{\mathds 1} +\half\,\nu\, {\boldsymbol{\sigma}}
\!\cdot\! {\mathbf n}_t,
\end{eqnarray}
where ${\mathbf n}_t =({\rm x}_t,{\rm y}_t,{\rm z}_t)$. Similarly
for the reference Hamiltonian $G$ we write
\begin{eqnarray}
G=\half\,v{\mathds 1} +\half\,\mu\, {\boldsymbol{\sigma}}
\!\cdot\!{\mathbf g}
\end{eqnarray}
for a unit vector ${\mathbf g}$. Bearing in mind the relations
\begin{eqnarray}
{\dot u}\propto {\tr} [H_t,X]=0\quad {\rm and} \quad [H_t,G]= \half
\ri \nu\mu\, {\boldsymbol{\sigma}}\!\cdot\!({\mathbf n}_t \times
{\mathbf g})
\end{eqnarray}
we find that (\ref{eq:1}) reduces to
\begin{eqnarray}
\frac{\rd {\mathbf n}_t}{\rd t} = \omega\, {\mathbf n}_t \times
({\mathbf n}_t \times {\mathbf g}),  \label{eq:2}
\end{eqnarray}
where $\omega=\lambda\nu\mu$. From
\begin{eqnarray}
\frac{\rd({\mathbf n}_t \cdot {\mathbf n}_t)}{{\rd t}} \propto
{\mathbf n}_t \cdot({\mathbf n}_t \times({\mathbf n}_t\times{\mathbf
g}))=0
\end{eqnarray}
we see that the norm of ${\mathbf n}_t$ remains constant under
(\ref{eq:2}). Without loss of generality we work with the basis in
which $G$ is diagonal, and choose ${\mathbf g}=(0,0,1)$. In terms of
the usual spherical parametrisation in the $G$-basis we have
${\mathbf n}_t =(\sin\theta_t \cos\phi_t,\sin\theta_t \sin\phi_t,
\cos\theta_t)$. Therefore, (\ref{eq:2}) reduces to:
\begin{eqnarray}
{\dot\theta}_t=\omega \sin\theta_t \quad {\rm and} \quad
{\dot\phi}_t=0. \label{eq:430}
\end{eqnarray}
Solving these, we obtain
\begin{eqnarray}
\cos\theta_t=\tanh(c_0- \omega t) \quad {\rm and} \quad
\phi_t=\phi_0, \label{eq:43}
\end{eqnarray}
where $c_0=\tanh^{-1}(\cos\theta_0)$ and $\phi_0$ are initial
values. The solution $H_t$ to (\ref{eq:1}) is thus
\begin{eqnarray}
H_t = \half\! \left(\! \begin{array}{cc} u_0\!-\!\nu \tanh(\omega
t-c_0) & \nu\,{\rm sech}(\omega t-c_0)\,\re^{-{\rm i} \phi_0} \\
\nu\,{\rm sech}(\omega t-c_0)\,\re^{{\rm i} \phi_0} & u_0 \!+\! \nu
\tanh( \omega t-c_0) \end{array} \!\right)\!.
\end{eqnarray}
A straightforward calculation shows that the eigenvalues of $H_t$
are time-independent, and that
\begin{eqnarray}
\lim_{t\to\infty} H_t = \half \left( \begin{array}{cc} u_0-\nu & 0
\\ 0 & u_0+\nu \end{array} \right). \label{eq:5}
\end{eqnarray}
Thus, the Hamiltonian is asymptotically diagonalised in the
$G$-basis. Observe that ${\tr}H_t$ and ${\det}H_t$ are conserved
quantities. Therefore, the flow induced by (\ref{eq:2}) for fixed
initial values $u_0$ and $|{\bf n}_0|$ is confined to a two-sphere
${\mathcal L}$, which can be identified with the state space of a
two-level system (i.e. the complex projective line). Since $u_0$ and
$|{\bf n}_0|$ are constant, in what follows we shall fix these two
variables and focus our attention upon the associated sphere
${\mathcal L}$ parameterised by the dynamical coordinates
$(\theta_t,\phi_t)$.

The fact that (\ref{eq:2}) defines a gradient flow
\begin{eqnarray}
\rd x^a = -\half\, \lambda\nu g^{ab}\nabla_b G(x) \rd t \label{eq:6}
\end{eqnarray}
on ${\mathcal L}$, where we use local coordinates
$(x^1,x^2)=(\theta,\phi)$ on the sphere, can be seen as follows.
First, in terms of these coordinates the inverse metric on the
sphere is
\begin{eqnarray}
g^{ab} = \frac{4}{\sin^2\theta} \left( \begin{array}{cc}
\sin^2\theta & 0 \\ 0 & 1 \end{array} \right). \label{eq:7}
\end{eqnarray}
We define a function $G(x)$ on the sphere ${\mathcal L}$ as follows:
\begin{eqnarray}
G(\theta,\phi)=\half(v+\mu\cos\theta).  \label{eq:7a}
\end{eqnarray}
This is obtained by taking the `expectation' of reference
Hamiltonian $G$ in a pure state corresponding to the point
$(\theta,\phi)$ on ${\mathcal L}$. Then, a short calculation using
(\ref{eq:6}), (\ref{eq:7}), and (\ref{eq:7a}) shows that the
dynamical equations (\ref{eq:430}) correspond to the gradient flow
(\ref{eq:6}).

\begin{figure}
\begin{center}
  \includegraphics[scale=0.5]{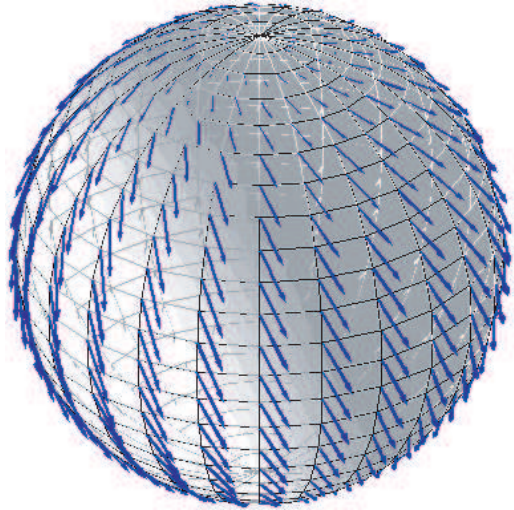}
  \caption{\textit{Flow on the sphere ${\mathcal L}$}. The
  vector field generated by the unitarily-modified gradient-flow
  (\ref{eq:8}) is plotted. The first term in (\ref{eq:8}) generates
  a rotation around the $G$-axis, while the second term generates
  geodesic flows toward the south pole. The axis ${\mathbf n}_0$ of
  the initial Hamiltonian $H_0$ spirals around the $G$-axis
  ${\mathbf g}$ and is asymptotically aligned with the latter.
  \label{fig:1}
  }
\end{center}
\end{figure}

We remark, incidentally, that the dynamical equation (\ref{eq:1})
can be modified to include a unitary term:
\begin{eqnarray}
\frac{\rd H_t}{\rd t} = -\ri[H_t,G] - \lambda \left[H_t,[H_t,G]
\right], \label{eq:8}
\end{eqnarray}
without greatly affecting its physical characteristics. In the
$2\times2$ example considered here, the only change occurs in the
phase, so that instead of $\phi_t=\phi_0$ we have $\phi_t
=\phi_0+\mu t$. Thus, according to (\ref{eq:1}) the eigenvectors of
$H_t$ evolve `straight' toward those of $G$ (i.e., along geodesics),
whereas under (\ref{eq:8}) they `spiral' toward those of $G$. This
is illustrated in Figure~\ref{fig:1} where we plot the vector field
on the sphere defined by the dynamical equation (\ref{eq:8}).

We also note that the diagonalisation property of the double-bracket
evolution (\ref{eq:1}) has been applied to the analysis of Toda
lattices \cite{ratiu}, dispersions in the Euler-Poincar\'e equations
\cite{bloch}, couplings in photorefractive media \cite{Brockett},
and flow equations in renormalisation group \cite{wegner}. Although
here we consider the case in which $G$ is fixed, it is possible to
vary $G$ in time (that is, vary the direction of ${\bf g}$). Then
the dynamical equation (\ref{eq:1}) can be used to characterise
\textit{quantum control} (cf.~\cite{helmke} for a related idea). In
this context it would be interesting to investigate the role of
\textit{geometric phases for observables}, when the control
Hamiltonian $G$ is varied along a loop in ${\mathcal L}$.

Having defined a natural gradient flow on the invariant 2-spheres
foliating the space of Hamiltonians, we now consider a dynamical
model on a given sphere ${\mathcal L}$ such  that an arbitrary
initial Hamiltonian $H_0$ evolves---according to
(\ref{eq:1})---towards the reference frame determined by $G$, but at
the same time  is randomly perturbed in all directions in ${\mathcal
L}$ by a pair of independent Brownsian motions. The dynamical model,
in particular, will possess the following properties: (i) the
eigenvalues of $H_t$ remain constant in time, and (ii) the
probability distribution over ${\mathcal L}$ evolves toward an
equilibrium distribution characterised by the standard canonical
density function. Although rather elaborate, this  model can be
treated analytically by identifying any given surface of the
foliation with the space of pure states of a two-level system, which
permits application of  the model for the thermalisation of quantum
states introduced in \cite{brody2}.

We consider first a stochastic differential equation of the form
\begin{eqnarray}
\rd x^a = \mu^a \rd t + \kappa \sigma^a_i \rd W_t^i \label{eq:sde}
\end{eqnarray}
on ${\mathcal L}$ (viewed as a real two-sphere). Here $\kappa$ is a
constant, the drift $\mu^a$ is a vector field on ${\mathcal L}$, and
the vectors $\{\sigma^{a}_{i}\}_{i=1,2}$ constitute an orthonormal
basis in the tangent space of ${\mathcal L}$ such that $g^{ab} =
\sigma^{a}_{i} \sigma^{b}_{j} \delta^{ij}$ and $\sigma^{a}_{i}
\sigma^{b}_{j}g_{ab}=\delta_{ij}$. We note that $\rd x^{a}$ is the
covariant Ito differential~\cite{hughston}, and that the standard
$2$-dimensional Wiener process $\{W^i_t\}$ satisfies $\rd W^{i}_{t}
\rd W^{j}_{t} = \delta^{ij}\rd t$. By straightforward calculation
one verifies  \cite{brody2} that the density function $\rho_t(x)$ on
${\mathcal L}$ associated with the stochastic evolution
(\ref{eq:sde}) satisfies the Fokker-Planck equation
\begin{eqnarray}
\frac{\partial}{\partial t}\,\rho_t(x) = -\nabla_{a}(\mu^{a}\rho_t)
+ \half \kappa^{2}\nabla^{2}\rho_t . \label{eq:11}
\end{eqnarray}
For our model we require that the drift vector $\mu^a$ represent the
double-bracket gradient flow (\ref{eq:1}). This is achieved by
choosing $\mu^a=-\frac{1}{2}\,\kappa^2 \lambda \nabla^a G$, where
$\kappa^2=\nu$. Then it follows from a theorem of
Zeeman~\cite{zeeman} that there exists a unique stationary solution
to (\ref{eq:11}), given by the canonical density
\begin{eqnarray}
\rho(x) = \frac{\exp(-\lambda G(x))}{\int_{\mathcal P}\exp(-\lambda
G(x))\rd V}. \label{eq:13}
\end{eqnarray}

To illustrate these results in more explicit terms we consider a
system consisting of a single spin-$\frac{1}{2}$ particle immersed
in an external magnetic field. The Hamiltonian is then $H=-{\mathbf
B} \cdot {\mathbf S}$, where ${\mathbf B}$ denotes the field and
${\mathbf S}$  the spin vector. The direction of the field ${\mathbf
B}$, however, is subject to fluctuations around its stable
direction, specified by $G$ (directed along the $z$-axis).
Calculating the orthonormal basis $\sigma^a_i$ on the sphere, we
obtain the stochastic equations for the variables $(\theta,\phi)$:
\begin{eqnarray}
\left\{ \begin{array}{l} \rm d\theta_t = \omega\, \sin\theta_t \rd t
+ \sqrt{2\nu}(\rd W_t^1+\rd W_t^2)
\\ \rd \phi_t = -\frac{1}{\sin\theta_t}\sqrt{2\nu}
(\rd W_t^1-\rd W_t^2).  \end{array} \right. \label{eq:14}
\end{eqnarray}
The associated Fokker-Planck equation reads
\begin{eqnarray}
{\dot\rho} = -\omega(\cos\theta + \sin\theta\,
\partial_\theta) \rho + 2 \nu{\textstyle{\left(
\partial_\theta^2\!+ \frac{1}{\sin^2\theta}
\partial_\phi^2 \right)}}\rho, \label{eq:15}
\end{eqnarray}
where $\partial_\theta=\partial/\partial\theta$ and $\partial_\phi
=\partial/\partial\phi$. The asymptotic solution is the following
canonical density function:
\begin{eqnarray}
\rho(\theta,\phi)= \frac{\lambda\mu}{2\pi\sinh(\frac{1}{2}
\lambda\mu)}\,\exp\left(-\half\lambda\mu\cos\theta\right).
\label{eq:16}
\end{eqnarray}
Direct substitution shows that (\ref{eq:16}) is the stationary
solution to (\ref{eq:15}). It follows from (\ref{eq:16}) and the use
of the spherical (Fubini-Study) volume element $\rd V=\frac{1}{4}
\sin\theta \rd\theta\rd\phi$ that the equilibrium mean Hamiltonian
is
\begin{eqnarray}
\langle H\rangle = \half \left( \begin{array}{cc} u_0+\nu\langle
\cos\theta\rangle_\lambda & 0 \\ 0 & u_0-\nu\langle
\cos\theta\rangle_\lambda, \end{array} \right), \label{eq:17}
\end{eqnarray}
where $\langle \cos\theta\rangle_\lambda = 2/\lambda\mu-
1/\tanh(\frac{1}{2} \lambda\mu)$. We may regard the parameter
$\lambda$ as representing the `inverse temperature' for the
Hamiltonian: if the noise level is high ($\lambda\ll1$), then the
direction of the external field ${\mathbf B}$ on the average lies
close to the $xy$-plane so that $\langle \cos\theta\rangle_\lambda
\simeq0$; whereas if the noise level is low  $\lambda\gg1$, then the
field ${\mathbf B}$ on the average is parallel to the $z$-axis and
we have $\langle \cos\theta\rangle_\lambda\simeq-1$.

\begin{figure}
\begin{center}
  \includegraphics[scale=0.5]{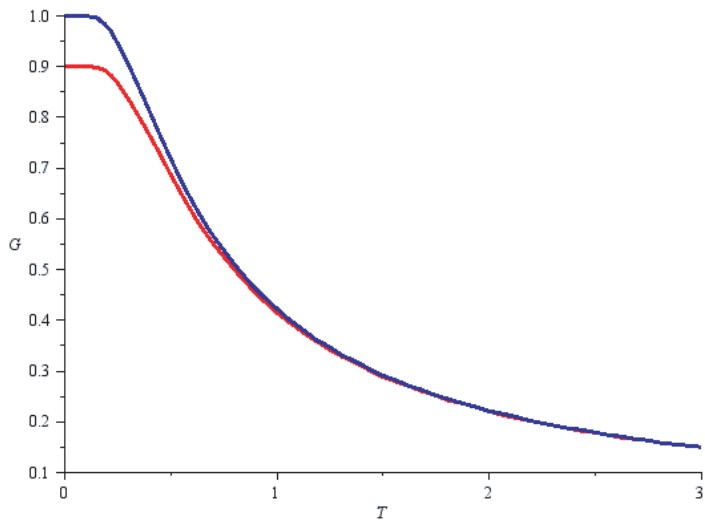}
  \caption{\textit{Quenched and annealed averages of $G$}. The
  functions $\langle G\rangle_Q$ and $\langle G\rangle_A$ are
  plotted against the temperature $T=\beta^{-1}$, where we set
  $\lambda^{-1}=0.1$, $v=0$, $\nu=1$, and $\mu=2$ so that
  $G=\sigma_z$.
  The `quenched magnetisation' $\langle\sigma_z\rangle_Q$ does not
  attain the maximum value $1.0$ at zero temperature unless
  $\lambda^{-1}=0$.
  \label{fig:2}
  }
\end{center}
\end{figure}

Let us now consider how the statistical theory of Hamiltonians
presented here can be applied to quantum statistical mechanics. In
this context it is natural to borrow ideas from the spin glass
literature~\cite{Virasoro}. We may take the averaged Hamiltonian
$\langle H\rangle_\lambda$ as the starting point of  the
analysis---this gives the analogue of an \textit{annealed} average.
In this regime the expectation of an observable $O$ is given by
\begin{eqnarray}
\langle O\rangle_A = \frac{{\tr} \left(O\re^{-\beta \langle
H\rangle_\lambda}\right)}{{\tr} \left(\re^{-\beta \langle
H\rangle_\lambda}\right)}.
\end{eqnarray}
Such an expectation, however, will involve the use of the averaged
Hamiltonian $\langle H\rangle_\lambda$ whose eigenvalues differ from
those of $H$. Alternatively, we may use the `unaveraged' Hamiltonian
to compute the thermal expectation of an observable $O$, regarded as
a function on a specified invariant surface in the above-described
foliation of the space of Hamiltonians, and then take its
average---this gives the analogue of a \textit{quenched} average:
\begin{eqnarray}
\langle O\rangle_Q = \left\langle \frac{{\tr} (O\re^{-\beta
H})}{{\tr} (\re^{-\beta H})}\right\rangle_\lambda.
\end{eqnarray}
A short calculation shows that the canonical quenched average of the
Hamiltonian $G$ is
\begin{eqnarray}
\langle G\rangle_Q = \half \mu\, {\textstyle \tanh\left(
\half\beta\nu\right) \left( \frac{1}{\tanh\left( \frac{1}{2}
\lambda\mu\right)}- \frac{2}{\lambda\mu} \right)},
\end{eqnarray}
whereas the canonical annealed average of $G$ is
\begin{eqnarray}
\langle G\rangle_A = \half \mu\, {\textstyle \tanh\left[
\half\beta\nu \left( \frac{1}{\tanh\left( \frac{1}{2}
\lambda\mu\right)}-\frac{2}{\lambda\mu} \right) \right]}.
\end{eqnarray}
These averages are plotted in Figure~\ref{fig:2}. These results
suggest a new line of studies on the extended quantum statistical
mechanics of disordered systems.

The explicit analysis presented above is for the most part confined
to $2\times2$ systems. In higher dimensions, the double-bracket
evolution equation (\ref{eq:1}) still defines an isospectral
gradient flow in the space of Hamiltonians. Thus, the procedure for
a statistical analysis of Hamiltonians as outlined above is
naturally extendable to higher dimensions. However, in higher
dimensions the equivalence of the Schr\"odinger and Heisenberg
picture for the nonunitary motion (\ref{eq:1}) breaks down (that is
to say, the generic surface foliating the space of Hamiltonians is
not isomorphic to the associated space of pure states). Instead, in
higher dimensions, the relevant foliation consists of certain
subspaces of higher-dimensional spheres. Nevertheless, there exist
unitary-invariant measures on these spaces, which can be used to
formulate the theory in an analogous manner. In particular, the
equilibrium state resulting from the thermalisation dynamics remains
canonical in the sense that it is proportional to the canonical
density $\exp(-\lambda\,{\tr}(GH))$ just as in the $2\times2$
example (cf. \cite{brockett2}). The remaining open problem is the
precise geometrical description of the relevant gradient flows in
higher dimensions, and the specification of  the associated measures
to calculate partition functions.

\vskip 10pt The authors thank A.~M.~Bloch, J.~Feinberg,
J.~E.~Marsden, B.~K.~Meister, T.~S.~Ratiu, and, in particular,
E.~J.~Brody and R.~Brockett, for comments and stimulating
discussions. DDH thanks the Royal Society of London for partial
support by its Wolfson Merit Award. 

\end{document}